\documentclass[aps,prb,superscriptaddress,twocolumn,amsmath,amssymb,notitlepage, longbibliography]{revtex4-2}
\usepackage{graphicx}% Include figure files
\usepackage{dcolumn}% Align table columns on decimal point
\usepackage{bm}% bold math
\usepackage{hyperref}% add hypertext capabilities
\usepackage{xcolor}

\def\beq{\begin{equation}}
\def\eeq{\end{equation}}

\begin{document}

%\preprint{APS/123-QED}

\title{Quantitative theory of magnetic properties of elemental praseodymium}% Force line breaks with \\

\author{Leonid V. Pourovskii}
% \altaffiliation[Also at ]{Ecole Polytechnique}%Lines break automatically or can be forced with \\
\email{leonid@cpht.polytechnique.fr}
\affiliation{CPHT, CNRS, \'Ecole polytechnique, Institut Polytechnique de Paris, 91120 Palaiseau, France
}%
\affiliation{Coll\`ege de France, Université PSL, 11 place Marcelin Berthelot, 75005 Paris, France}


\author{Alena Vishina}
\affiliation{
 Uppsala University, Department of Physics and Astronomy\\
 Box 516, SE-75120, Uppsala, Sweden
}
\author{Olle Eriksson}
\affiliation{
 Uppsala University, Department of Physics and Astronomy\\
 Box 516, SE-75120, Uppsala, Sweden
}
\affiliation{WISE-Wallenberg Inititative Materials Science, Uppsala University, Box 516, SE-751 20 Uppsala, Sweden}
%\affiliation{
\author{Mikhail I. Katsnelson}
\affiliation{%
 Institute for Molecules and Materials, Radboud University, Heijendaalseweg 135, 6525AJ Nijmegen, The Netherlands\\
}
\affiliation{WISE-Wallenberg Inititative Materials Science, Uppsala University, Box 516, SE-751 20 Uppsala, Sweden}

\date{\today}% 

\begin{abstract}
Elemental Pr metal crystallizes in the double hexagonal close packed (dhcp) structure and is unique among rare-earth elements in featuring a localized partially filled 4$f$ shell without ordered magnetism. Experimental evidence attributes this absence of magnetism to a singlet crystal-field (CF) ground state of the Pr 4$f^2$ configuration, which is energetically well isolated from excited magnetic doublets. Here, we construct a realistic effective magnetic Hamiltonian for dhcp Pr, by combining density-functional theory with dynamical mean-field theory, in the quasiatomic Hubbard-I approximation. Our calculations fully determine the CF potential and predict singlet CF ground states at both inequivalent sites of the dhcp lattice. The intersite exchange interactions, obtained from the magnetic force theorem, are found to be insufficient to close the CF gap to the magnetic doublets. Hence, ab-initio theory is demonstrated to explain the unusual, non-magnetic state of elemental Pr. Extending this analysis to the (0001) surface of Pr, we find that the singlet ground state remains robust preventing conventional magnetic orders. Nevertheless, the gap between the ground state and the lowest excited singlet is significantly reduced at the surface, opening the possibility for exotic two-dimensional multipolar orders to emerge within this two-singlet manifold.
\end{abstract}

\maketitle

%\tableofcontents

%\section{\label{sec:level1}Introduction}
The interest in the 17 rare-earth (RE) elements as vital components to functional materials is steadily increasing \cite{WorldMin}. RE-containing modern materials find their applications in devices that convert mechanical energy to electricity, in fuel cells and in batteries \cite{Skomski1999,LIN2024}. They are also vital components in light-emitting diodes, in LCD screens and in lasers \cite{WANG2015,Li2022,Song2013,QUAZI2016,STECKL2007}. Chemically, in solids, they are often found in a trivalent electronic configuration \cite{Johansson1984,Johansson1979} where the outermost valence electrons form itinerant states that contribute to the chemical bonding. The electronic states of the 4{\it f} shell of the lanthanides (a subgroup of the rare-earths involving 15 elements) have limited ability to hybridize with other states (except La, Ce and Yb). Instead, these electron states form a localized magnetic moment that has significant spin and orbital contributions (as summarized excellently in Ref.~\cite{Jensen1991}), that for the most part can be understood from Russel-Saunders (RS) coupling. This involves a treatment of the angular momenta (spin $S$, orbital $L$, and total $J$) of an $f^n$ configuration ({\it n}-electrons in the {\it f}-shell) that is atomic-like. From this, one would expect that any lanthanide element with $14>${\it n}$>$0 should form a significant magnetic moment, which is true for the most of the lanthanides and forms an essential background to, e.g., functional magnetic materials like Nd$_2$Fe$_{14}$B~\cite{Coey1996}.

Surprisingly, for the solid phase of elemental Pr, that is trivalent with an $f^2$ electronic configuration in elemental form, the expected magnetic state (with $S$=1, $L$=5 and $J$=4) is completely missing~\cite{Bleany1963,Lebech1971}. Instead, Pr is a temperature-independent paramagnet. This paramagnetic ground state can be forced to undergo a meta-magnetic transition at extremely high applied magnetic field~\cite{Jensen1991}. Furthermore, at millikelvin temperatures the nuclear moments can order via nuclear RKKY interaction~\cite{McEwen1981,Jensen1991}, but the 4{\it f} shell, with two electrons that in RS coupling are expected to form a stable spin-paired state, shows no experimental evidence of magnetism. 

On a theoretical model level, the enigmatic, non-magnetic state of Pr has been somewhat explained by crystal field (CF) theory. In this model, one assumes that Heisenberg exchange of a possible magnetic state results in a total energy that is larger than that of a CF split non-magnetic, singlet level of the $J$=4 angular momentum state. A singlet CF state has indeed been detected %experimentally 
by inelastic neutron scattering (INS) for the hexagonal site of the dhcp lattice that elemental Pr solidifies in \cite{Houmann1975,Houmann1979}. Combined with a singlet state that is inferred as a possible state of the cubic site~\cite{Lea1962,Houmann1979}, the absence of magnetic order of Pr has been proposed. 

Unfortunately, there is no solid experimental evidence of CF level splittings of Pr atoms at the cubic site of the dhcp structure~\cite{Houmann1979,Jensen1991}. A theory that does not rely on experimental data of the CF levels of Pr is also missing, both for the cubic and hexagonal sites of dhcp Pr. This means that the total singlet state of elemental Pr is still undetected, with full evidence missing both from theory and from experiment. In addition, the theory of the interatomic exchange interaction of elemental dhcp Pr, e.g. 
%as calculated 
based on the magnetic force theorem~\cite{Liechtenstein1984,Liechtenstein1987} (for a review of this method see Ref.~\cite{Szilva2023}), is missing, primarily due to the complexity of the electronic structure of the 4{\it f} shell of the rare-earths. There have been some attempts to extract information about the interatomic exchange by fitting experimentally observed magnetic excitations together with susceptibility and magnetization curves data. Long-range RKKY interactions together with a significant two-ion anisotropy needed to be included into the effective magnetic Hamiltonian for quantitative agreement with experimental energies of low-energy excitation branches \cite{Houmann1979}.
%(Leonid please add here how this was done),
However, the long-range nature of RKKY interactions, together with the limited number of excitation branches accessible to INS render such fits quite uncertain. 
%extremely large amount of spin-wave energies are needed for this calculation, one must conclude that such estimates have so far been less reliable. 

Based on the discussion above, one must conclude that an accurate theoretical calculation or experimental estimate of the energetics relevant to determining the energy balance and the competition between CF splitting of a non-magnetic spin-singlet state and a magnetically ordered state of dhcp Pr is missing. This paper attempts to provide such an analysis for bulk dhcp Pr, based on ab-initio electronic structure theory combined with dynamical mean field theory. In addition, we address a possible magnetic state of Pr for both surface terminations with a cubic and hexagonal site.

%4.	Summarized excellently in {\it Rare Earth Magnetism, Structures and Excitations}, J.Jensen and A.R.Mackintosh (Clarendon Press, Oxford 1991).

%5.	J.M.D. Coey, {\it Rare-earth iron permanent magnets 1996}.

%6.	CF paper on expt. for Pr, maybe Leonid can help here

%7.	M.T. Hutchings, J. Solid State Physics {\bf 16}, 227 (1964).

%8.	K. R. Lea, M. J. M. Leask and W. P. Wolf, J. Phys. Chem. Solids {\bf 23}, 1381 (1962).

%9. A.I.Liechtenstein, M. I. Katsnelson, and V. A. Gubanov, Journal of Physics F: Metal Physics {\bf 14}, L125 (1984); A.I.Liechtenstein, M. I. Katsnelson, V. P. Antropov, and
%V. A. Gubanov, Journal of Magnetism and Magnetic Materials {\bf 67}, 65 (1987).

%10. A.Szilva et al., Rev. Mod. Phys. {\bf 95}, 035004 (2023).

\section{\label{sec:model}Theoretical model}

A minimal model considered here is an effective low-energy Hamiltonian comprising CF and intersite Heisenberg exchange, that act within the $^3H_4$ ground-state multiplet (GSM) of the Pr$^{3+}$ ion:
\beq\label{eq:Heff}
\hat{H}_{\mathrm{eff}}=\sum_i \hat{H}_{i}^{\mathrm{CF}}-\sum_{ij} I_{ij}\mathbf{J}_i\mathbf{J}_j, 
\eeq
where $H_{i}^{\mathrm{CF}}$ is the  single-site CF term  acting on the site $i$, $I_{ij}$ is the intersite exchange coupling between the total angular momentum operators $J_ {i(j)}^{\alpha}$ ($\alpha$=$x,y,z$) at the corresponding sites.
$\mathbf{J}_{i(j)}$ is the  Cartesian vector of these operators (in contrast to 
a classical treatment of the angular momentum that 
%of these operators 
considers $\mathbf{J}_{i(j)}$ to be an ordinary Cartesian vector).
%A classical treatment of the angular momentum of these operators considers $\mathbf{J}_{i(j)}$ to be the Cartesian vector instead of  these operators. \LP{The previous sentence is unclear, if $J^{\alpha}$ are operators, what means "classical" in this case?} \textbf{I tried to fix, please check - MK} 
In Eqn.~\ref{eq:Heff} we neglect the anisotropic exchange coupling and magnetoelastic effects that were shown to be important for a detailed description of field-induced magnetic response and the excitation spectra of Pr~\cite{Houmann1979,Jensen1991}. These smaller terms are unlikely to strongly affect the overall energetics and the competition between a magnetic ordered state and a possible singlet CF state.

In the standard Stevens formalism, the CF term reads
\beq\label{eq:H_CF}
\hat{H}^{\mathrm{CF}}=\sum_{kq}B_k^qO_k^q,
\eeq
where $O_k^q$ are the Stevens operator (Ref.~\cite{Stevens1952}, see, e.~g. Refs.~\cite{Hutchings1964,Kuzmin2007} for a review) of the rank $k$ and projection $q$ acting within the  $^3H_4$ multiplet, $B_k^q$ are the corresponding CF parameters (CFPs). Due to the point group symmetry of the hexagonal and cubic sites in the dhcp lattice, the CFPs can be non-zero for the following $kq$ combinations only:
\begin{flalign}
\mathrm{hex:} \quad kq & = 20, 40, 60, 66  \\
\mathrm{cub:} \quad kq & = 20, 40, 43,  60, 63, 66. \label{eq:CF_cub}
\end{flalign}
In this work, we use first-principles approaches to calculate all parameters in Eqn.~\ref{eq:Heff}. 
%In the following subsections, we describe our approach for calculating the CFPs $B_k^q$  and the exchange interactions $I_{ij}$
We detail  approach for calculating the CFPs $B_k^q$  and the exchange interactions $I_{ij}$ in the Methods.

\section{\label{sec:CF}Crystal field and intersite exchange in bulk Pr}

\subsection{\label{sec:CF_levels} Crystal-field parameters and levels}\label{sec:CFP_bulk}

The calculated CFPs for the two dhcp sites are listed in Table~\ref{tab:CF_pars}, where we also show the CFPs values of Ref.~\cite{Houmann1979} estimated from neutron scattering and magnetization measurements. For the hexagonal site, our values overall agree with the experimental estimates within the uncertainties of the latter. For the cubic site, only two CFPs were experimentally estimated from fitting magnetic susceptibility and the lowest magnetic excitation  \cite{Houmann1979}; our calculated  full set of CFPs differ significantly from those estimates, with the previously neglected $B_4^3$, $B_6^3$ , and $B_6^6$ terms taking large values. We note that the Stevens operators $O_k^q$ are not normalized to unity. Their norm (which can be defined as, e.~g., $\sqrt{\mathrm{Tr} [O_k^q\cdot O_k^q]}$) scales approximately as $J^k$.  
%In particular, the contributions due to the  $B_4^3$, $B_6^3$, and $B_6^6$ CFPs at the cubic site are more significant than that of the $B_2^0$ CFP.  
Hence, the formally small $B_k^q$ CFPs for $k=4$ and 6 in  Table~\ref{tab:CF_pars} lead to significant contributions to the CF potential.

\begin{figure*}[!tb]
 \centering
 \includegraphics[width=2\columnwidth]{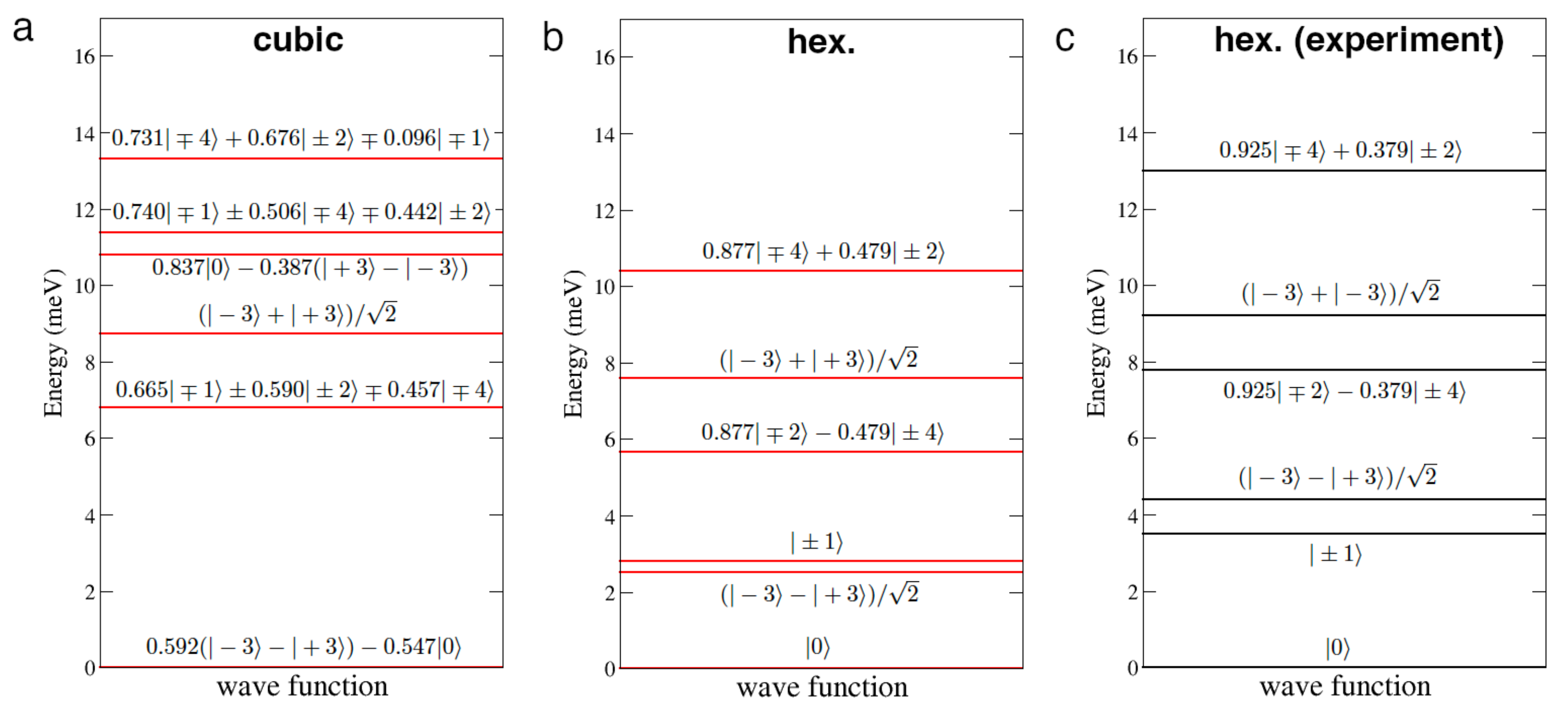}
\caption{Calculated crystal-field splitting of the Pr $^3H_4$ configuration for the cubic (a) and hexagonal (b) site in bulk dhcp Pr. The CF wavefunctions are written  in the $|M\rangle\equiv |J=4;M_J\rangle$ basis and are defined in the same coordination frame as the CFPs in Table~\ref{tab:CF_pars}. In panel (c), we reproduce the experimentally inferred CF level scheme of Ref.~\cite{Houmann1979} for the hexagonal site.}
\label{fig:CF_lev}
\end{figure*}

The ``cubic'' site of the dhcp-structure possesses the actual cubic symmetry only in the case of an ideal $c/a=$3.266 ratio ($c/a$=3.222 in dhcp Pr). In the case of ideal $c/a$, the CFP $B_2^0$ becomes zero, while $B_k^q$ for $k=4, 6$ and $q > 0$ are related to the corresponding $B_k^0$ by well-known relations~\cite{Hutchings1964}. We evaluated the ``ideal cubic'' CFP set starting from the values of $B_4^0$ and $B_6^0$ calculated for the actual cubic site. The resulting ``ideal'' values of $B_4^3$ and $B_6^6$ are underestimated by about 30\% as compared to their actual values in Table~\ref{tab:CF_pars}. 
Large values of high-rank CFPs $B_k^q$ for non-zero $q$  have been previously found using the present method~\cite{Delange2017,Pourovskii2020} as well as experimentally~\cite{Passos2023} in  rare-earth intermetalics $R$Co$_5$, where they were shown to arise due 4$f$ hybridization with conduction states~\cite{Pourovskii2020}. The same hybridization contribution to the CFPs and, thus, appears to be important also in the case of dhcp Pr. 

In Figs.~\ref{fig:CF_lev}a and b, we display the calculated CF level schemes for the two sites, expressing the CF wavefunctions as superpositions of the $J_z$ eigenstates $|J=4;M_J\rangle$ of the $^3H_4$ atomic configuration. The CF ground state is a singlet in both cases, with the gap to the lowest excited doublet $|\pm 1\rangle$ for the hexagonal and cubic sites is 2.8 and 6.8~meV, respectively, as compared to the experimental estimates of 3.5 and 8.4~meV \cite{Houmann1979}. The theoretical values are thus underestimated by about 20\%. We note that the singlet $(|-3\rangle-|+3\rangle)/\sqrt{2}$  is lower than the $|\pm 1\rangle$ by 0.3~meV (Fig.~\ref{fig:CF_lev}b), in contrast to the experimental CF scheme for the hexagonal site (Fig.~\ref{fig:CF_lev}c), though the excitations from the GS singlet to this excited one will not be directly detectable by inelastic neutron scattering experiments. Otherwise, the order of theoretical CF levels and the composition of corresponding wavefunctions agree with the experimental picture, apart from a systematic underestimation by about 20\%. 

Evaluating the cubic-site level scheme from  the ideal-cubic CFPs described above, we find the overall splitting reduced by about 2~meV. The second (doublet) and third (singlet) levels in Fig.~\ref{fig:CF_lev}a merge into a triplet, as well as the fourth (singlet) and fifth  (doublet) levels also merging into the second triplet.

\begin{table*}[tp]
	\begin{center}
		\begin{ruledtabular}
                \renewcommand{\arraystretch}{1.3}
			\begin{tabular}{l c c c c c c c}             
                  \multicolumn{7}{c}{\textbf{Bulk}} \\
                  %\hline           
                  & $B_2^0 \times 10^2$ & $B_4^0 \times 10^4$ & $B_4^3\times 10^4$ & $B_6^0\times 10^4$ & $B_6^3 \times 10^4$ & $B_6^6\times 10^4$  \\
                  \hline
			Hex. site & 14.0  & -4.17  &  -  & 0.82  & -  & 10.3  \\
                Hex. site (exp. estimate) & 19$\pm$4 & -5.7$\pm$5 &  - & 1.0$\pm$0.1 & - & 9.6 \\
                     Cubic site & 3.05 & 11.6  & -462  & 0.9 & 10.0  & 11.2  \\
                     Cubic site (exp. estimate) &  & 29   &   & 0.8 &   &   \\
                  \hline\hline
                 \multicolumn{7}{c}{\textbf{(0001) surface}} \\
                 \multicolumn{7}{c}{Hexagonal termination} \\
                 surface layer (hex.) & -2.26 & -6.17 & -15.07 & 0.97 & 3.06 & 4.20 \\
                 subsurface layer (cub.) & -3.93 & 8.06 & -182.09 & 0.81 & 13.45 & 8.81 \\
                 \hline
                 \multicolumn{7}{c}{Cubic termination} \\
                 surface layer (cub.) & -5.76 & 2.48 & 81.1 & 1.1 & 5.89 & 3.92 \\
                 subsurface layer (hex.) & 4.88 & -3.96 & 141 & 0.86 & 6.68 & 7.61 \\
                  %\hline     
			\end{tabular}
		\end{ruledtabular}
	\end{center}
	\caption{Calculated CF parameters for bulk and (0001) relaxed surface of dhcp Pr (in meV). The bulk values are  compared to the corresponding experimental estimates of Houmann {\it et al.}~\cite{Houmann1979}.  Only two CF parameters were estimated in Ref.~\cite{Houmann1979} for the cubic site. We use the coordinate frame with $y||b$, $z||c$.}
   \label{tab:CF_pars} 
\end{table*}

\subsection{\label{sec:Jij} Intersite exchange interactions}
 
\begin{figure}[h]
 \centering
 \includegraphics[scale=0.38]{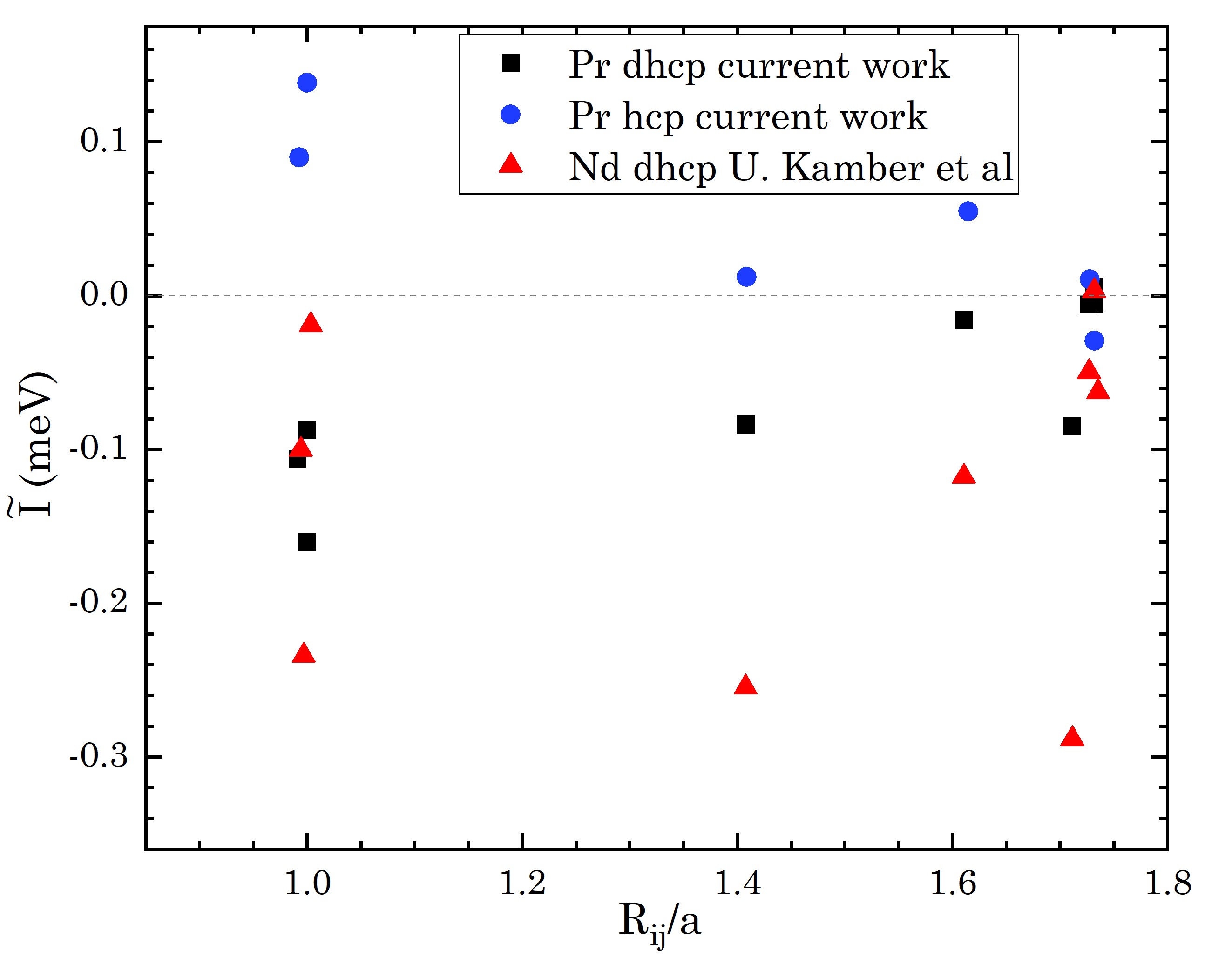}
\caption{Exchange interactions $\tilde I_{ij}$ for the dhcp crystal structures of Pr (current work) and Nd \cite{Kamber2020}, as well as for the hcp Pr (current work)}
\label{fig:dhcpJij}
\end{figure}

The exchange interactions (see Eqns.~\ref{eq:Heff2a} and \ref{eq:Heff2} for their definition) were calculated both for the dhcp and a hypothetical hcp structure, similar to what was done in Ref.~\cite{Kamber2020}. 
%PAW-PBE Pr\_3 pseudopotential was used, the plane wave basis energy cut-off was 500 eV with the k-point grid of 11$\times$11$\times$11. 
In Fig.~\ref{fig:dhcpJij}, we show the exchange interactions of dhcp Pr and compare them to those of dhcp Nd, reported in Ref.\cite{Kamber2020}. The intersite exchange of hcp Pr is also added for comparison. We can see that $\tilde I_{ij}$s in dhcp Pr are similar to that of dhcp Nd \cite{Kamber2020}, as at short range the interactions are weak and antiferromagnetic.  Pr, in general, has weaker exchange interactions compared to Nd. Similarly to Nd, we find that Pr has stronger exchange interactions in the hcp structure. hcp $\tilde I_{ij}$s are ferromagnetic for short-range coupling for both materials, additional details can be found in {\it Supplementary Section 1}.

 \begin{figure}[!t]
 \centering
 \includegraphics[width=0.9\columnwidth]{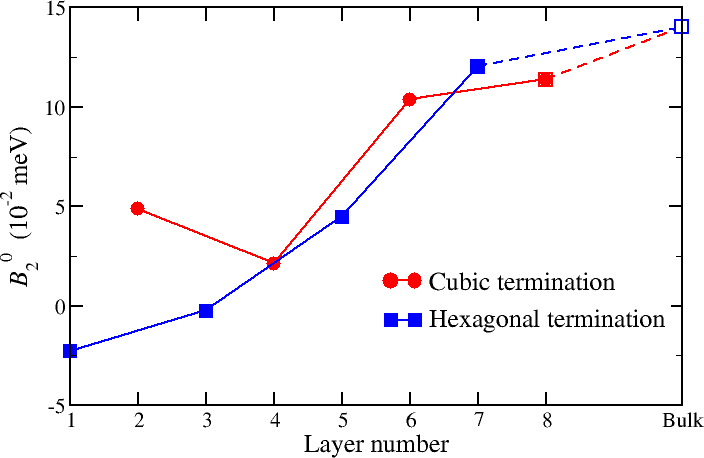}
 \includegraphics[width=0.9\columnwidth]{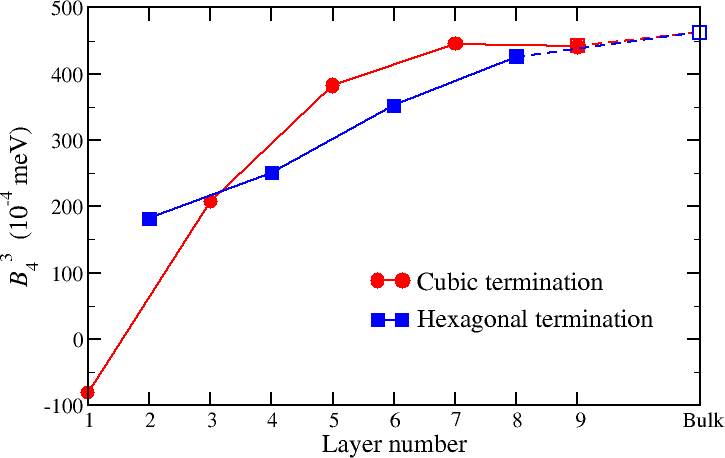}
\caption{Evolution of the $B_2^0$ CFP on hexagonal sites (top panel) and the $B_4^3$ CFP on cubic sites (bottom panel) vs the layer depth with respect to the surface. The surface layer is the first one; the corresponding values for bulk are shown on the right-hand side. Since the sign of $B_4^3$ in the global coordination frame is flipped between the two dhcp cubic sites related by the inversion symmetry, we display the value  $B_4^3$ in its local frame (i.~e., we flip the sign of $B_4^3$ for every second cubic layer starting from the bulk one).}
\label{fig:CFPs_surf}
\end{figure}

\begin{figure*}[!ht]
 \centering
 \includegraphics[width=2\columnwidth]{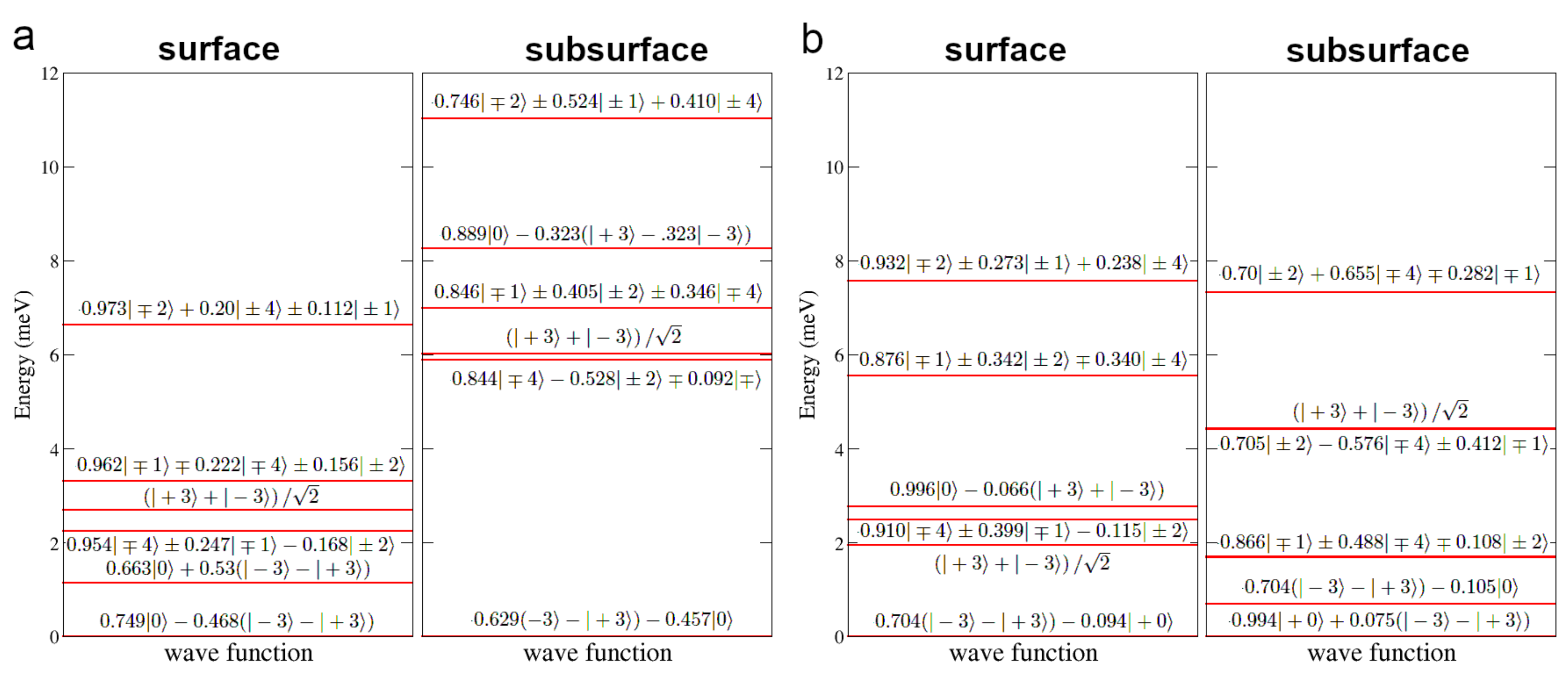}
\caption{Calculated crystal-field level splitting of the Pr $^3H_4$ multiplet at the (0001) dhcp surfaces with hexagonal (a) and cubic (b) termination. The CF wavefunction representation and coordination frame are the same as in Fig.~\ref{fig:CF_lev}. For both cases, we show the levels for the surface and subsurface site. In the subsurface layer the site symmetries are reversed with respect to the surface one, becoming cubic in (a) and hexagonal in (b), respectively.}
\label{fig:CF_surf}
\end{figure*}

\section{\label{sec:surf}Crystal field and intersite exchange on the (0001) surface}

\subsection{Intersite exchange interactions}\label{sec:surfexchange}

In order to investigate the difference in exchange interactions between the bulk and surface layers of the dhcp structure of Pr, we performed slab calculations for the two possible surface terminations (with a cubic or a hexagonal layer as the surface layer). An {\it ab initio} optimization of the slab geometry was first performed, 
%using VASP, 
for details see the {\it Methods} section.
%An optimization of the slab geometry using VASP was performed for 11 layers with a vacuum region of 15 {\AA}, where the three central layers were fixed to their bulk positions. 
As was similarly shown in Ref.\cite{Kamber2020} for dhcp Nd, the bulk-like magnetism is restored within just a few atomic layers. 
We found that for the hexagonal termination, the surface layer relaxes 2.4\% inward, while the subsurface layer moves by 4.8\% outward. For the cubic termination, these values are 1.34\% and 4.4\%.

Magnetic moments were then calculated with RSPt with similar results for the surface layers of both terminations and the bulk. The 1$^{st}$ (surface) layer it is 1.77 $\mu_{B}$ and 1.78 $\mu_{B}$ for the cubic and hexagonal terminations, respectively. For the 2$^{nd}$ (subsurface) layer, the values are 1.72 $\mu_{B}$. In the case of the bulk, we obtained 1.72 $\mu_{B}$ and 1.71 $\mu_{B}$ for the cubic and hexagonal sites. For the surfaces, the exchange parameters were calculated with the RSPt code in a manner similar to the bulk case using the relaxed slab structure. %The simulations were done using 5600 k-points of the full BZ. 
The values of $\tilde I_{ij}$ for the surface and subsurface layers deviate considerably from the bulk data, as is outlined in detail in the {\it Supplementary Section} 2. As an example, $\tilde I_{1,2}$ is FM (0.14 meV for the cubic termination and 0.02 meV for the hexagonal termination), where bulk values are AFM, as we see in Fig.~\ref{fig:dhcpJij}. At the 3$^{rd}$ layer, the values of $\tilde I_{ij}$s are almost restored to the bulk values.

\subsection{Crystal field}\label{sec:surfCF}

We calculated the CFPs for both surface terminations of the (0001) dhcp surface using the relaxed DFT geometries described above. The relaxed interlayer distances were employed for the first four surface layers; the distances for deeper layers were fixed at the bulk value.  
The calculated CFPs values for both terminations  are listed in Table~\ref{tab:CF_pars}. For the hexagonal sites, the on-site inversion symmetry is lifted by the surface, with their point group symmetry $\bar{6}m2$ correspondingly reduced to $3m$. The CFPs $B_4^3$ and $B_6^3$ thus become non-zero in hexagonal layers. With our choice of the supercells, the global inversion symmetry with respect to the middle-layer cubic sites is preserved. Therefore, all sites have the same set (\ref{eq:CF_cub})  of real non-zero CFPs.

For both terminations, one observes CFPs to strongly deviate from the corresponding bulk values that are also displayed in Table~\ref{tab:CF_pars}. In particular, the values of $B_2^0$ for hexagonal surface and subsurface layers are significantly reduced compared to the bulk, while $B_4^3$ takes rather large values. One also observes a significant overall reduction of high-rank CFPs for the cubic site. As noted in Sec.~\ref{sec:CFP_bulk},  high-rank CFPs are enhanced by hybridization effects; one can expect those effects to be reduced on the surface due to the reduced coordination number. 
In Fig.~\ref{fig:CFPs_surf} we show the evolution of  the $B_2^0$ CFP for the  hexagonal sites and the $B_4^3$ CFP for the cubic ones vs. layer's depth with respect to the surface.   The value of $B_4^3$ is seen to increase quite monotonously and in a similar way for both terminations; the bulk value is virtually reached at the fifth layer. The behavior of $B_2^0$ is broadly similar, though displaying some oscillations vs. the depth.

The calculated CF level schemes for surface and subsurface layers are shown in Fig.~\ref{fig:CF_surf}. As could be anticipated from the reduced CFPs values, we find correspondingly reduced CF splittings  at the surface. The lifting of inversion symmetry on the hexagonal sites leads to mixing of the bulk GS ($|0\rangle$) with  the excited singlet  $\left(|-3\rangle-|+3\rangle\right)/\sqrt{2}$ by $B_k^3$ CF terms. For the hexagonal surface layer, a strong  reduction of $B_2^0$ diminishes the splitting between those levels to about 1~meV. Similarly, one finds the splitting between the singlet GS and excited levels to decrease significantly for the cubic termination surface layer due to the decay of all high-rank CFPs  $B_k^q$ for $q>0$ (Table~\ref{tab:CF_pars}). For the subsurface (hexagonal) layer of the same termination, we find the CF gap between the GS singlet and the excited one to reduce to 0.7~meV only. This can be explained by a significantly enlarged interlayer distance between  the subsurface and second surface layer predicted by our DFT calculations, which results in an especially strong deviation from the hexagonal symmetry reflected by a large value of $B_4^3$ (Table ~\ref{tab:CF_pars}). Indeed, setting this CFP to zero, which  mimics approximately restoring hexagonal symmetry, increases the gap back to above 2~meV.

\section{\label{sec:MF_bulk} Magnetic vs nonmagnetic ground state}

Having calculated the CFPs and intersite exchange interactions $\tilde{I}_{ij}$, we obtain the full effective Hamiltonian, $\hat{H}_{\mathrm{eff}}$, in Eqn.\ref{eq:Heff}  for dhcp Pr. This allows for a full determination of the magnetic ground state. We solved the Hamiltonian, using the single site quantum mean field approach of the McPhase package~\cite{Rotter2004} together with an in-house module that implements the $J_{\alpha}$ and $O_k^q$ operators. We converted the calculated inter-site exchange coupling of the classical model, Eqn.~\ref{eq:Heff2a}, to the $J=4$ quantum model of Eq.~\ref{eq:Heff} as $I_{ij}=\tilde{I}_{ij}/20$, 
taking into account the length of quantum angular momentum as $\sqrt{J(J+1)}$. 

These results are discussed below for the bulk and surfaces of dhcp Pr.

\subsection{Bulk}

Solving $\hat{H}_{\mathrm{eff}}$ (Eqn.\ref{eq:Heff}) for dhcp phase we correctly obtained a nonmagnetic state, with both crystallographic sites having the same singlet ground state as shown in Fig.~\ref{fig:CF_lev}. It agrees with experimental observations and illustrates the competition between interatomic exchange, where the latter favors a magnetically ordered state, while crystal field effects for Pr favor a non-magnetic, singlet state. Following the experimental observations, the singlet state has the lowest energy. It means that the energy gain that would come from a magnetically ordered state, as quantified by the second term of Eqn.~\ref{eq:Heff}, is smaller than the gain of the singlet crystal field effect that arises due to the Coulombic interaction of the $J=$4 state of Pr in the dhcp crystal structure. In the Supplementary Section 3, we analyze the magnetic contribution to the specific heat, and a Schottky anomaly that arose due to the CF levels of Pr. Also presented in the {\it Supplementary Section 4} are the field-induced magnetism and comparison to experimental data. There, in particular, a metamagnetic transition observed at 30 T is absent on this level of theory. As discussed in the Supplementary, other terms of the Hamiltonian than the ones discussed in Eqn.~1 may be needed for a full description of the field-induced magnetism of dhcp Pr. 

%Olle suggests to move the text below to the appendix
%Furthermore, in Fig.~\ref{fig:Cp} we show the calculated magnetic contribution to the specific heat. Note that it exhibits no discontinuities that would indicate phase transitions; instead one notices only a Schottky anomaly peaked at about 30~K, apparently generated by the lowest-energy CF excited levels of the hexagonal site.

\subsection{Surface}

Similarly to the Hamiltonian for bulk Pr, we calculated the magnetic state of the dhcp Pr (0001) surface, for both terminations (cubic and hexagonal) using the layer-resolved CFPs and $I_{ij}$'s described above. The mean-field simulation supercell employed 15 and 17 layers for the hexagonal and cubic terminations, respectively. %, followed by a vacuum spacer of sufficient thickness to fully decouple supercells along the $c$ axis. 
In both cases, we found the singlet paramagnetic state with no magnetic phases appearing. The decrease of CF splitting at the surface (Fig.~\ref{fig:CF_surf}) is thus not sufficient for the magnetic order to prevail. We note that even when increasing the values of the surface $I_{ij}$'s by a factor of two does not affect this result, showing that the non-magnetic GS state is sufficiently robust with respect to small changes in the effective Hamiltonian (\ref{eq:Heff}).

 \begin{figure}[!htb]
 \centering
 \includegraphics[width=1\columnwidth]{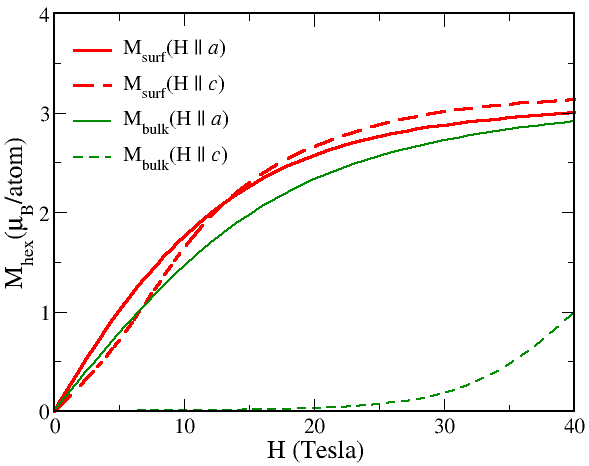}
\caption{Magnetic moment of the hexagonal site in the bulk and in the surface layer vs. applied field.}
\label{fig:Mhex}
\end{figure}

In Fig.~\ref{fig:Mhex} we display the magnetization of the hexagonal surface layer vs. external field applied along the in-plane ($a$) and out-of-plane ($c$) directions compared to the corresponding bulk data. The field in these calculations was applied to all layers of the supercell to simulate an experimental uniform field. One observes a drastic reduction of the anisotropy on the surface, with the magnetization curves for both directions almost coinciding, in a sharp contrast to the bulk case. The origin of this anisotropy collapse can be understood from the CF level scheme for the  hexagonal surface layer (Fig.~\ref{fig:CF_surf}a). The singlet $\left(|-3\rangle+|+3\rangle\right)/\sqrt{2}$ is shifted down by about 5~meV  as compared to the bulk (Fig.~\ref{fig:CF_lev}b) and occurs right above the magnetic doublet. Moreover, due to the $B_k^3$ CFPs being non-zero on the surface (Table~\ref{tab:CF_pars}), the singlet GS also acquires a $|\pm3\rangle$ admixture.  In the bulk, only the in-plane moment operators $M_{x(y)}=g_JJ^{x(y)}$ couple the singlet GS $|0\rangle$ to other CF levels, namely, to the low-lying CF doublet $|\pm1\rangle$. In the surface case,  one may easily show that the matrix elements of the out-of-plane magnetic moment operator $M_z$ between the GS singlet and the lowest excited singlet are also non-zero. As a result, $M_{\alpha}$ for all three directions couple the GS with excited CF levels located at about the same energy, leading to an isotropic behavior in the applied field. 

A strongly reduced CF gap between the GS and excited singlet levels could be naively expected to lead to a magnetic ordering on the (0001) surface.  However, both these levels are superpositions of the $|0\rangle$ and  $\left(|-3\rangle-|+3\rangle\right)/\sqrt{2}$ states. All matrix elements of the total angular momentum operators $J^{\alpha}$ can be easily shown to be zero within the space spanned by those two states. Hence, conventional magnetic orders cannot be hosted by them. The splitting between the singlet GS and the lowest magnetic doublet remains about 2~meV.

\section{Discussion and Conclusions}

Using a Hamiltonian (\ref{eq:Heff}) that consists of CF splittings and Heisenberg inter-site exchange terms, with all parameters obtained from ab initio, electronic structure theory, we demonstrate the absence of magnetic order in bulk, dhcp Pr. 
This result is in perfect agreement with experimental observations. It is important to note that the level of electronic structure theory used here goes beyond standard formulations based on density functional theory and the common approximations of the exchange and correlation functional. Instead, we used the combination of density functional with dynamical mean field theory in the Hubbard-I approximation for the calculation of crystal field parameters, since it is difficult to see how theories that treat electron correlations less precisely, could have the needed accuracy to reproduce the singlet state of dhcp Pr. 
Hence, electronic structure theory, as outlined here, is capable of reproducing the complex magnetic state of bulk Pr and this is a central result obtained in this investigation.  

The theoretical results of the Pr (0001) surface are in some ways similar to those of bulk Pr. As shown in Sec.~\ref{sec:surfCF}, the splitting between the GS singlet and lowest magnetic doublet on the hexagonal sites is only moderately reduced at the surface or subsurface hexagonal layers. In contrast, our calculations predict the splitting between the GS and lowest excited singlet to reduce for surface and subsurface hexagonal layers down to 1~meV and below (Fig~\ref{fig:CF_surf}).
The space of the two singlets cannot host conventional dipolar magnetic moments. However, higher-order multipolar operators can have non-zero matrix elements in this space. Multipolar orders and fluctuations arising from two closely spaced non-magnetic singlets have been discussed, e.g., for the prototypical hidden-order system URu$_2$Si$_2$ \cite{Haule2009} and the unconventional superconductor UTe$_2$ \cite{Khmelevskyi2023}. 

Evaluating inter-site multipolar exchange in Pr is beyond the scope of the present work.  However, we have explored possible multipolar orders on the basis of the compositions of the two singlets. To that end, we evaluated matrix elements of the multipolar Stevens operators $O_k^q$ (see, e.~g.,  Ref.~\cite{Rudowicz2004} for the explicit form of odd-$k$ Stevens operators) within the space of the two singlets finding non-zero matrix elements only for the even-rank (charge) multipoles with $q=0,3$ and the odd-rank (magnetic) multipoles with $q=-3$. Since the surface CF Hamiltonian, eq.~(\ref{eq:H_CF}) and Table~\ref{tab:CF_pars}, already contains all even-rank terms with $q=0,3$, no even-rank order parameter is possible within the two-singlet space. Among odd-$k$ multipoles $O_k^{-3}$, we find the largest matrix elements for the octupolar moment operator,  
%are cubic polynomials of $J^{\alpha}$ and can thus be active  in the space of two singlets. Specifically, we find non-zero matrix elements  within this space for the octupole 
$O_3^{-3} =\frac{i}{2}\left[\left(J^{-}\right)^3-\left(J^{+}\right)^3\right]$. A sufficiently strong exchange coupling between these octupoles can lead to an unconventional surface octupolar order on the Pr (0001) surface. 

Direct detection of high-rank magnetic multipolar orders represents a significant challenge even when they occur in the bulk, see, e.g., Refs.~\cite{Santini2009,Maharaj2020,Paramekanti2020}. Moreover, the relevant spectroscopical x-ray or neutron scattering probes are typically bulk sensitive. However, octupolar orders can manifest themselves indirectly by coupling to the strain~\cite{Patri2019,Ye2023}.
Namely, an appropriate strain applied to an octupole-ordered system can generate conventional dipole moments that are easily detectable. In the context of surface science, strains can be generated by 
depositing adatoms or by the inclusion of surface impurities. We simulated the effect of strain in the presence of an octupolar $O_3^{-3}$order by supplementing the single-site CF Hamiltonian with the corresponding exchange and strain-induced terms
\beq\label{eq:H_o}
\hat{H}^{\mathrm{CF}}+I_{\mathrm{o}}O_{3}^{-3}+KO_{2}^q,
\eeq
where $I_o$ is the exchange mean-field due to the  $O_3^{-3}$ order; the last term is induced by strain coupled to the quadrupole operator $O_{2}^q$ of matching symmetry. By diagonalizing (\ref{eq:H_o}) using the CFPs for hexagonal surface and subsurface layers (Table~\ref{tab:CF_pars}) we obtained the GS expectation value for dipole magnetic moments induced in the octupolar phase by various strains. We find the largest effect in the case of $x^2-y^2$ orthogonal strain. This strain, which couples to the $O_2^{2}$ quadrupole,  corresponds to the hexagonal layer being compressed along the $a$ lattice parameter and extended in the orthogonal direction.  The strain is found to induce an in-plane dipole magnetic moment along $a$. % Leonid can we give a values here like '..upto a value of xx muB, for strains of order 3 \%' ? 
Therefore, the conjectured octupole order could indeed be detectable by inducing local strains,  e.~g., by depositing adatoms. The dipole moments are then expected to appear in the vicinity of the lattice perturbation below the octupolar transition temperature.

\section*{Methods}

\subsection{\label{sec:level2} Calculation of crystal-field parameters}

In order to calculate the CFPs, we use the approach of Ref.~\cite{Delange2017} that is based on  density-functional theory (DFT)+dynamical mean-field theory~\cite{Georges1996} (DMFT) framework~\cite{Anisimov1997_1,Lichtenstein1998,Kotliar2006} in the Hubbard-I (HI) approximation \cite{Hubbard1963,Lichtenstein1998}.  Our charge-density self-consistent implementation~\cite{Aichhorn2009,Aichhorn2011}  of DFT+HI is based on the Wien2k linearized augmented-plane-wave (LAPW)   full-potential code~\cite{Wien2k} and "TRIQS" library for implementing DMFT~\cite{Parcollet2015,Aichhorn2016}.  In these calculations, we used LDA exchange correlations and the LAPW basis cutoff $R_{\mathrm{mt}}K_{\mathrm{MAX}}$=8. The on-site Coulomb repulsion was specified by the parameters $F^0=U=6$~eV and Hund's rule coupling  $J_H=0.7$~eV, which are in the commonly accepted range of Coulomb interaction parameters for Pr$^{3+}$ \cite{Carnall1989,Locht2016}. Moreover, the  CFPs calculated by DFT+HI  have been shown to be rather insensitive to varying $U$ and $J_H$ \cite{Delange2017}. The double-counting correction was calculated in the fully localized limit using the nominal occupancy $f^2$ of Pr, as has been shown to be appropriate for the  DFT+HI framework~\cite{Pourovskii2007}. We employ projective Wannier functions \cite{Aichhorn2009} to represent the 4{\it f} orbitals, see further details below. The experimental dhcp Pr lattice parameters $a=$3.672~\AA\ and $c=$11.833~\AA\ \cite{Jensen1991}  are used in bulk calculations. 

%In this study, we also consider a possible magnetic state of the surface of Pr. In particular, 
In order to evaluate the CF splitting at the (0001) surface, we carried out DFT+HI calculations using 17(15)-layer slabs in the case of cubic (hexagonal) terminations, respectively, with about 15~\AA\ of vacuum spacer. We always kept the in-plane lattice parameter at the bulk value, while the interlayer spacings were fixed at the relaxed DFT values obtained as  detailed below.

As shown in Refs.~\cite{Brooks1997,Delange2017}, the non-spherical 4{\it f}-electron charge density induces  a DFT self-interaction contribution to the CFPs. In order to suppress this unphysical contribution, we average the Boltzmann weights of the nine states within the $^3H_4$ GSM during self-consistent DFT+HI iterations, following the prescription of Ref.~\cite{Delange2017}. Another important issue in the DFT+HI CF calculations is the choice of the projective window to construct 4$f$ Wannier orbitals. By using a narrow window enclosing mainly the 4{\it f} bands, one may effectively take into account the impact of hybridization on the CFPs within the quasi-atomic HI approximation~\cite{Delange2017,Pourovskii2020}.  In the case of entangled 4{\it f} bands in rare-earth elemental metals, the choice of this window is not unambiguous, since the 4{\it f} manifold is not separated from other bands. We employ the projective window [-1.1:1.1]~eV around the centerweight of Kohn-Sham Pr-4$f$ band (which corresponds to [-0.86:1.34]~eV with respect to the Kohn-Sham Fermi level)  enclosing about 97\% of 4$f$-electron spectral weight. In the surface calculation we employ, with respect to the KS Fermi level,  the same projective window.

Having converged the DFT+HI calculations, we extract the CFPs for each crystallographically inequivalent Pr site $a$ from the 4$f$ one-electron level positions that read
\beq\label{eq:H_a}
H_a=E_a^0+H_a^{\mathrm{SO}}+H_a^{\mathrm{CF}},
\eeq
where the terms on the right-hand side are the uniform shift, spin-orbit interaction and crystal-field term.
We find inter-multiplet mixing in dhcp Pr to be totally negligible. Therefore, we calculated the matrix elements of $H_a$ in the basis of $J_z$ eigenstates  $|J;M\rangle \equiv|M\rangle$ of the Pr $f^2$ ground state multiplet $^3H_4$ and extracted the CFPs by fitting the resulting 
 matrix $\langle M|H_a|M'\rangle$ to the form (\ref{eq:H_CF}).
%see Refs.~\cite{Delange2017,Pourovskii2020} for more details on this procedure.

\subsection{\label{sec:Jij_cal}Calculation of Heisenberg exchange}

The calculations of the Heisenberg exchange were performed according to Refs.\cite{Liechtenstein1984,Liechtenstein1987,Szilva2023}. We considered the spin-exchange only to the Heisenberg interaction and therefore made the substitution
\beq\label{eq:Heff2a}
\sum_{ij} I_{ij}\mathbf{J}_i\mathbf{J}_j \to \sum_{ij} I_{ij}^{\prime}\mathbf{S}_i\mathbf{S}_j,
\eeq
where according to the Russel-Sunders coupling $S=J(g_J-1)$, where $g_J$ is the Land\'e $g$-factor. %\LP{Then $I_{ij}$ in the RHS is not the same, as in the LHS, i.~e. $I_{ij}\to I_{ij}/(g_J-1)^2$?} \textbf{Shall we use different letters for I's e.g. with and without hat or prime? - MK}  
Moreover, according to Refs.~\cite{Liechtenstein1984,Liechtenstein1987,Szilva2023}, we included the value of the spin angular momentum in the exchange parameter (see Eq.~1.3 in Ref.~14), which means that for the Heisenberg exchange of Eq.~\ref{eq:Heff} the following substitution was considered:

\beq\label{eq:Heff2}
\sum_{ij} I_{ij}^{\prime}\mathbf{S}_i\mathbf{S}_j \to \sum_{ij} \tilde I_{ij}\mathbf{e}_i\mathbf{e}_j .
\eeq
Note that these equations represent a system where exchange interaction is dominated by the spin moment, and not the orbital magnetic moment. Since it is primarily the itinerant valence electrons of the rare-earths that mediate the exchange for which the orbital moment is essentially quenched, the approach presented here captures the dominating contributions to the magnetic couplings.

For the calculations of $\tilde I_{ij}$, the full-potential linear muffin-tin orbital method (FP-LMTO) as implemented in the RSPt code \cite{Wills1987,Wills2010} was used. The PBE functional \cite{PBE} for exchange and correlation was employed. Experimental unit cell parameters were used for the calculations. 4{\it f} electrons were treated as localized and unhybridized particles with a magnetic moment according to Russel-Saunders coupling, according to the standard model of the lanthanides. The calculations were performed using 8000 {\it k}-points for bulk dhcp Pr.

To obtain the relaxed unit cell of the hcp structure, the Vienna Ab Initio Simulation Package (VASP) \cite{Kresse1993,Kresse1996} was used within the Projector Augmented Wave (PAW) method \cite{Blochl1994}. PAW-PBE Pr\_3 pseudopotential was employed, the plane wave basis energy cut-off was 500 eV with the k-point grid of 11$\times$11$\times$11. The optimization of the slab geometries was performed for 11 Pr layers with a vacuum region of 15 {\AA}, where the three central layers were fixed to their bulk positions. For the surfaces, the exchange parameters were calculated with the RSPt code in a manner similar to the bulk case using the relaxed slab structure. The simulations were done using 5600 k-points of the full BZ. 
%As was shown in Ref.\cite{Kamber2020}, the bulk-like magnetism is restored within just a few atomic layers. 

%\subsection{\label{sec:slab_relax} Slab structure relaxation}

%To calculate the exchange interactions for the surface layers, a slab geometry optimisation with 11 atomic layers was performed using
%Vienna Ab Initio Simulation Package (VASP) \cite{Kresse1993,Kresse1996} within the Projector Augmented Wave (PAW) method \cite{Blochl1994}. A vacuum region of 15 {\AA} was introduced. Atomic positions were relaxed using force minimisation for the four top and four bottom layers, keeping the inner three layers fixed to their bulk coordinates.

\begin{acknowledgments}
M.I.K and O.E. acknowledge support from the Wallenberg Initiative Materials Science for Sustainability (WISE) funded by the Knut and Alice Wallenberg Foundation (KAW) and the European Research Council through the ERC Synergy Grant 854843-FASTCORR. O.E. also acknowledges support from STandUPP, eSSENCE, the Swedish Research Council (VR) and the Knut and Alice Wallenberg Foundation (KAW-Scholar program) and NL-ECO: Netherlands Initiative for Energy-Efficient Computing (with project number NWA. 1389.20.140) of the NWA research program.  
The computations/data handling were enabled by resources provided by the National Academic Infrastructure for Supercomputing in Sweden (NAISS) (projects NAISS2024-5-75, NAISS2024-5-427, and NAISS 2024/1-18), partially funded by the Swedish Research Council through grant agreement no. 2022-06725.
 L. V. P. is thankful to the CPHT computer team
for support.
\end{acknowledgments}


\end{document}

\title{Supplementary Information for "Quantitative theory of magnetic properties of elemental praseodymium".}% Force line breaks with \\

\author{Leonid V. Pourovskii}
% \altaffiliation[Also at ]{Ecole Polytechnique}%Lines break automatically or can be forced with \\
\email{leonid@cpht.polytechnique.fr}
\affiliation{CPHT, CNRS, \'Ecole polytechnique, Institut Polytechnique de Paris, 91120 Palaiseau, France
}%
\affiliation{Coll\`ege de France, Université PSL, 11 place Marcelin Berthelot, 75005 Paris, France}


\author{Alena Vishina}
\affiliation{
 Uppsala University, Department of Physics and Astronomy\\
 Box 516, SE-75120, Uppsala, Sweden
}
\author{Olle Eriksson}
\affiliation{
 Uppsala University, Department of Physics and Astronomy\\
 Box 516, SE-75120, Uppsala, Sweden
}
\affiliation{WISE-Wallenberg Inititative Materials Science, Uppsala University, Box 516, SE-751 20 Uppsala, Sweden}
%\affiliation{
\author{Mikhail I. Katsnelson}
\affiliation{%
 Institute for Molecules and Materials, Radboud University, Heijendaalseweg 135, 6525AJ Nijmegen, The Netherlands\\
}
\affiliation{WISE-Wallenberg Inititative Materials Science, Uppsala University, Box 516, SE-751 20 Uppsala, Sweden}

\date{\today}	
\maketitle

\section{\label{app:subsec1} Interatomic exchange of hcp structures}

%\subsection{\label{app:subsec1} hcp structures}

\begin{figure}[h]
 \centering
 \includegraphics[width=0.6\columnwidth]{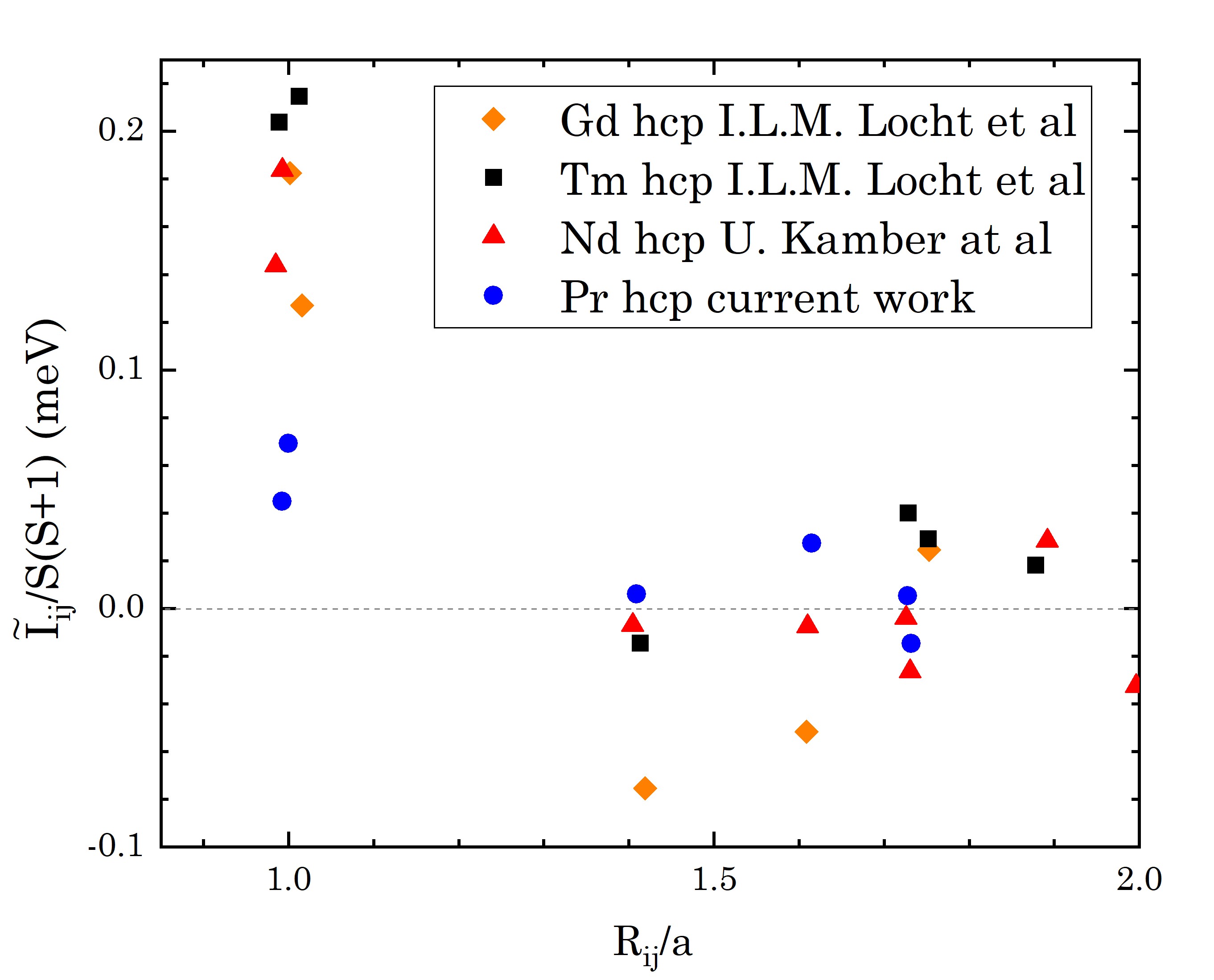}
\caption{Exchange interactions $\tilde I_{ij}/S(S+1)$ for the hcp structures of Pr (current work) compared to the values obtained for hcp Nd \cite{Kamber2020}, hcp Gd \cite{Locht2016}, and hcp Tm \cite{Locht2016}.}
\label{fig:hcpJij}
\end{figure}

In Fig.~\ref{fig:hcpJij} we present the $\tilde I_{ij}$ parameters  %divided by the value of the 4{\it f} spin moment 
for hypothetical hcp Pr obtained in current work as well as the exchange parameters  calculated in Ref.~\cite{Kamber2020} for hcp Nd, and in Ref~\cite{Locht2016} for hcp Gd and hcp Tm. 
In Fig.~\ref{fig:hcpJij} we divide  $\tilde I_{ij}$ by $S(S+1)$ to account for the length of quantum spins, see Eqns.~7 and 8 of the main text.
In the hcp structure, for all rare-earth elements analysed here, we observe a qualitatively similar behaviour with ferromagnetic nearest-neighbour interactions changing into oscillating values at larger distances, as reported previously for other lanthanides. 
 Quantitatively, $\tilde I_{ij}/S(S+1)$  differ significantly among the calculated RE hcp metals, reflecting the sensitivity of the RE intersite exchange to changes in the lattice parameters.

%You can use a subsection or subsubsection in an appendix. Note the
%numbering: we are now in Appendix~\ref{app:subsec}.

%Olle suggests to move the text below to the appendix

\section{\label{app:subsec2} Interatomic exchange of the surface of the dhcp structure}

\begin{figure*}[h]
 \centering
 \includegraphics[width=\columnwidth]{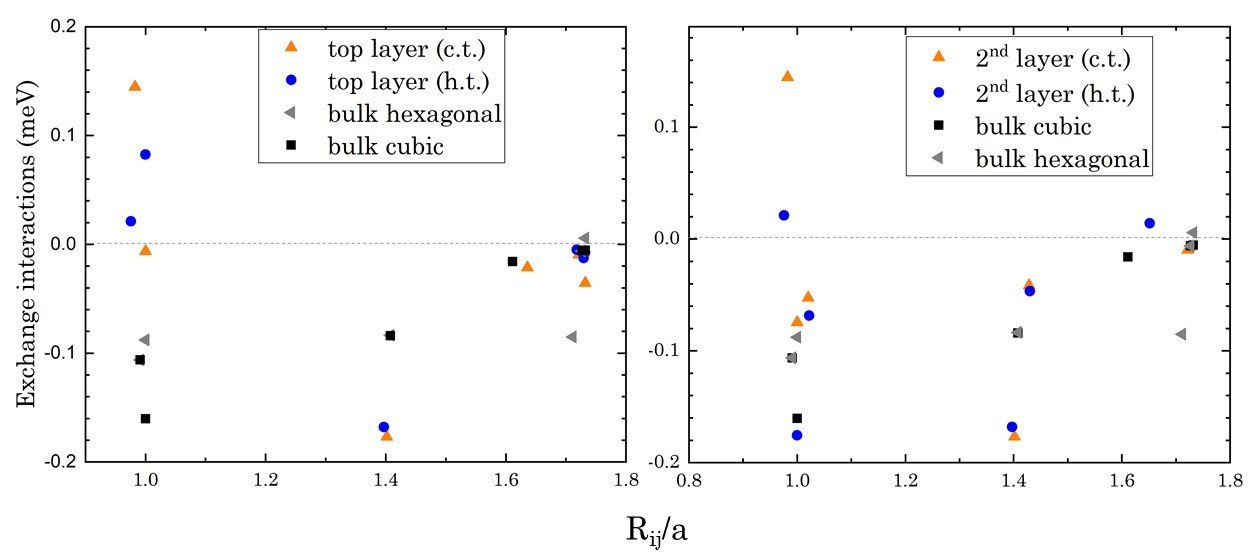}
\caption{(Color online) Exchange interactions $\tilde I_{ij}$ for the surface and the 2nd layer of Pr dhcp slab with hexagonal termination (h.t.) and cubic termination (c.t.).}
\label{fig:surfJij}
\end{figure*}

In Suppl. Fig.~\ref{fig:surfJij} we show the calculated interatomic exchange of dhcp Pr, as a function of distance between atom pairs. In the left part of the figure, we illustrate how the exchange interactions of the surface atoms compare to those of the bulk dhcp Pr. The right part of the figure shows the corresponding data for the subsurface atoms. Note that both surface and sub-surface atoms have a distinctly different interatomic exchange compared to bulk values. For the top layer, the $\tilde I_{ij}$ are positive and ferromagnetic, moving closer to the antiferromagnetic exchange of the bulk as we move further away from the surface. 

%\subsection{\label{sec:Jij} Intersite exchange interactions}

\section{\label{app:subsec3} Calculated magnetic specific heat}

\begin{figure}[htb]
 \centering
 \includegraphics[width=0.7\columnwidth]{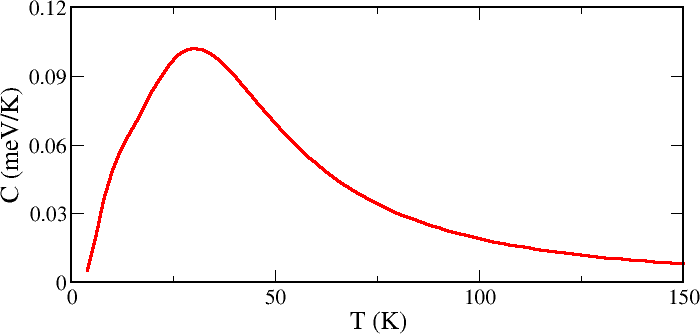}
\caption{ Calculated magnetic contribution to the specific heat (per atom) of dhcp Pr.}
\label{fig:Cp}
\end{figure}

In Suppl. Fig.~\ref{fig:Cp} we show the calculated magnetic contribution to the specific heat. Note that it exhibits no discontinuities that would indicate phase transitions; instead one notices only a Schottky anomaly peaked at about 30~K, apparently generated by the lowest-energy CF excited levels of the hexagonal site.
 
\section{\label{app:subsec4} Magnetisation of bulk, dhcp Pr}

\begin{figure}[htb]
 \centering
 \includegraphics[width=0.7\columnwidth]{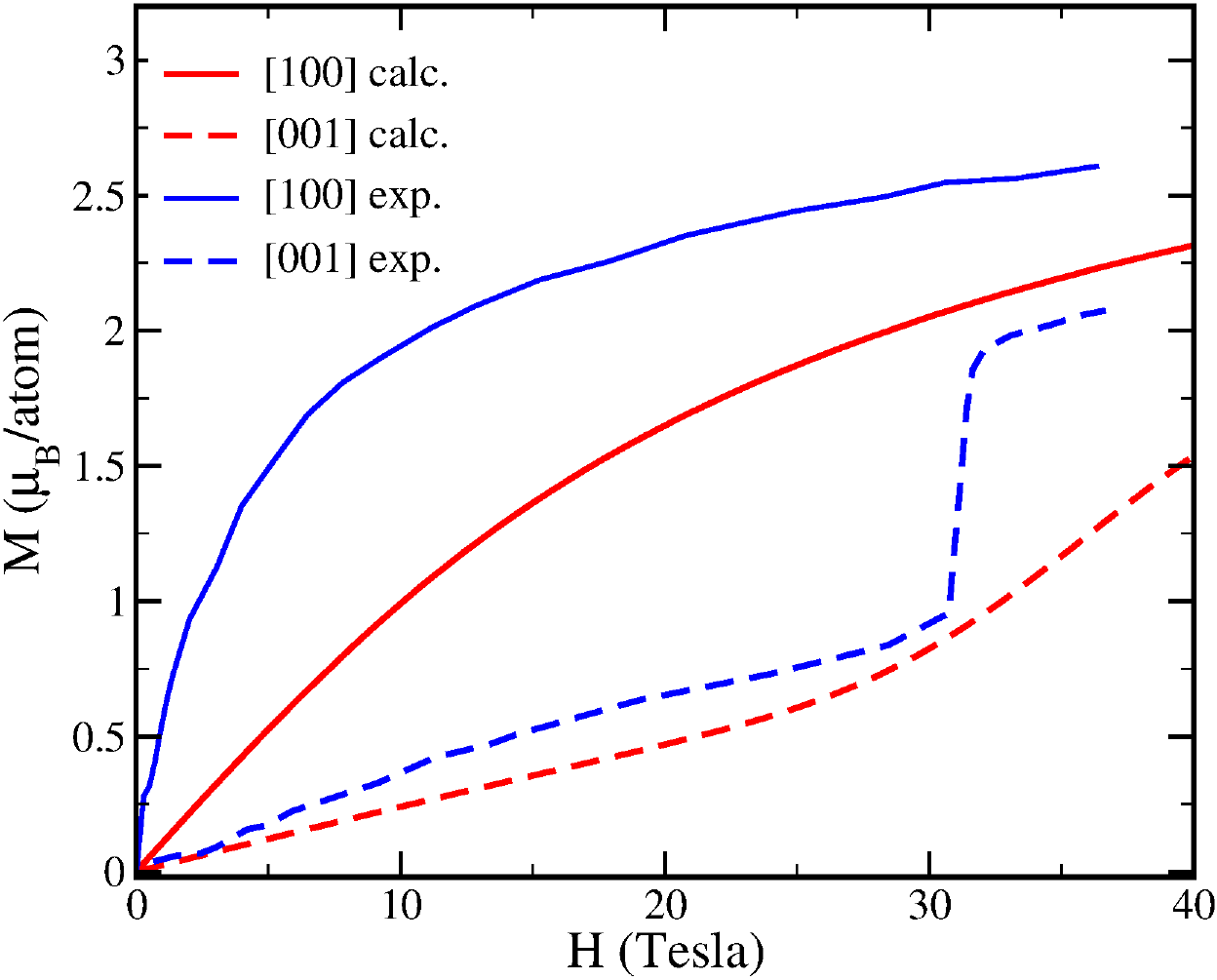}
\caption{Calculated magnetization curves in dhcp Pr (in red) compared to experiment (in blue). The experimental data are from Fig.~7.13 of Ref.~\cite{Jensen1991}.}
\label{fig:Mag_curves}
\end{figure}

We show the calculated magnetization curves for bulk Pr with a magnetic field applied along the [100] ($a$-axis) and [001] ($c$-axis) directions. This was obtained by supplementing the effective Hamiltonian, eqn.~1 of the main text,  with a Zeeman term. The resulting curves are shown together with the corresponding experimental data~\cite{Jensen1991} in Suppl. Fig.~\ref{fig:Mag_curves}. The easy-plane anisotropy is reproduced by the theory. However, theory on this level of approximation does not obtain a quantitative agreement with the measurements, with the anisotropy underestimated and the observed first-order metamagnetic transition at 30 Tesla absent. We notice that the suggested Hamiltonian~\cite{Houmann1979,Jensen1991} based on experimental data, includes significant two-site anisotropy as well as magnetoeslastic terms, which are neglected in the present theory. Those terms seem to be necessary to quantitatively reproduce the magnetization curves of dhcp Pr.

